\documentclass[aps,pra,amssymb,superscriptaddress,twocolumn]{revtex4-1}
\usepackage[pdftex]{graphicx}
\usepackage[dvipsnames]{xcolor}
\usepackage{bm}
\usepackage{amsmath}
\usepackage{ascmac}
\usepackage{color}
\usepackage{setspace}
\usepackage{amssymb}
\usepackage{latexsym}
\usepackage{mathtools}
\usepackage{lipsum}
\usepackage{braket}
\usepackage[colorlinks,linkcolor=RoyalBlue,citecolor=RoyalBlue,urlcolor=RoyalBlue]{hyperref}
\usepackage{MnSymbol}

\begin{document}

\title{Random Matrix Statistics in Propagating Correlation Fronts of Fermions} 

\author{Kazuya Fujimoto}
\affiliation{Department of Physics, Tokyo Institute of Technology, 2-12-1 Ookayama, Meguro-ku, Tokyo 152-8551, Japan}

\author{Tomohiro Sasamoto}
\affiliation{Department of Physics, Tokyo Institute of Technology, 2-12-1 Ookayama, Meguro-ku, Tokyo 152-8551, Japan}

\date{\today}

\begin{abstract}
We theoretically study propagating correlation fronts in non-interacting fermions on a one-dimensional lattice starting from an alternating state, where the fermions occupy every other site. We find that, in the long-time asymptotic regime, all the moments of dynamical fluctuations around the correlation fronts are described by the universal correlation functions of Gaussian orthogonal and symplectic random matrices at the soft edge. Our finding here sheds light on a hitherto unknown connection between random matrix theory and correlation propagation in quantum dynamics. 
\end{abstract}

\maketitle
{\it  Introduction.\text{-}} 
Propagation of correlation has been one of the central topics in quantum many-particle systems \cite{Eisert2015,Abanin2019,LRB2}. It is often associated with emergence of propagating correlation fronts, forming light-cone structures \cite{Burrell_2007,LCex1,LCex2,LCex3,Hauke2013,Bonnes2014,Carleo2014,Bertini2018,takasu2020}.  A fundamental result on the subject is the existence of a universal bound for a two-point correlator, known as the Lieb-Robinson bound found in 1972 \cite{Lieb1972}. Since then, correlation-front dynamics has attracted considerable attention, and the universal aspects have been extensively explored from various viewpoints such as entanglement entropy \cite{EE1,EE2,EE3,EE4} and operator spreading \cite{OP1,OP2,OP3,OP4}. Currently, the state-of-the-art experiments of cold atoms have observed the light-cone structures \cite{LCex1,LCex2,LCex3}. 

Appearance of propagating fronts is not necessarily restricted to correlation dynamics. For instance, particle distribution can develop into propagating particle-density fronts when the particles are initially distributed in a spatially restricted region. One of the typical situations is a one-dimensional system with a domain-wall initial condition, for which previous works \cite{Antal_1999, ogata2002, Hunyadi2004,Platini2005, Platini2007, Antal_2008, Jesenko_2011, Eisler_2013, Sabetta_2013, Viti_2016, Ljubotina2017, Stephan_2017, Misguich_2017, Misguich_2019, Collura_2018, Moriya_2019, Jin_2021, KPZ1,GHD1,GHD2,GHD3,GHD4,GHD5} have investigated the fundamental properties, e.g., the propagating speed of the front and the variance of the integrated particle current. In particular, Eisler and R\'acz theoretically studied dynamical fluctuation of the particle-density around the front of non-interacting fermions \cite{Eisler_2013}, finding that the dynamical fluctuation is characterized by universal eigenvalue distributions of the Gaussian Unitary Ensemble (GUE) \cite{mehta1,forrester2010} of random matrix theory. Subsequent works \cite{Viti_2016, Collura_2018, Moriya_2019, Jin_2021} have studied the details from various perspectives, such as dependence of initial states and effects of interactions.

Such dynamical fluctuation, however, has yet to be explored well in correlation-front dynamics. So far, previous literature on propagating correlation fronts has focused mainly on two-point or four-point correlators and has studied the fundamental aspects in terms of the Lieb-Robinson bound and the light-cone structures \cite{Eisert2015,Abanin2019,LRB2,Burrell_2007,LCex1,LCex2,LCex3,Hauke2013,Bonnes2014,Carleo2014,Bertini2018,takasu2020,Lieb1972,Bravyi2006,Chen2019,Tran2020,Kuwahara2020}. With this background, it is intriguing to explore universal nature of correlation dynamics beyond the conventional light-cone structures captured by low order correlators, by focusing on the dynamical higher-order fluctuation of correlation.

\begin{figure}[b]
\begin{center}
\includegraphics[keepaspectratio, width=8.5cm]{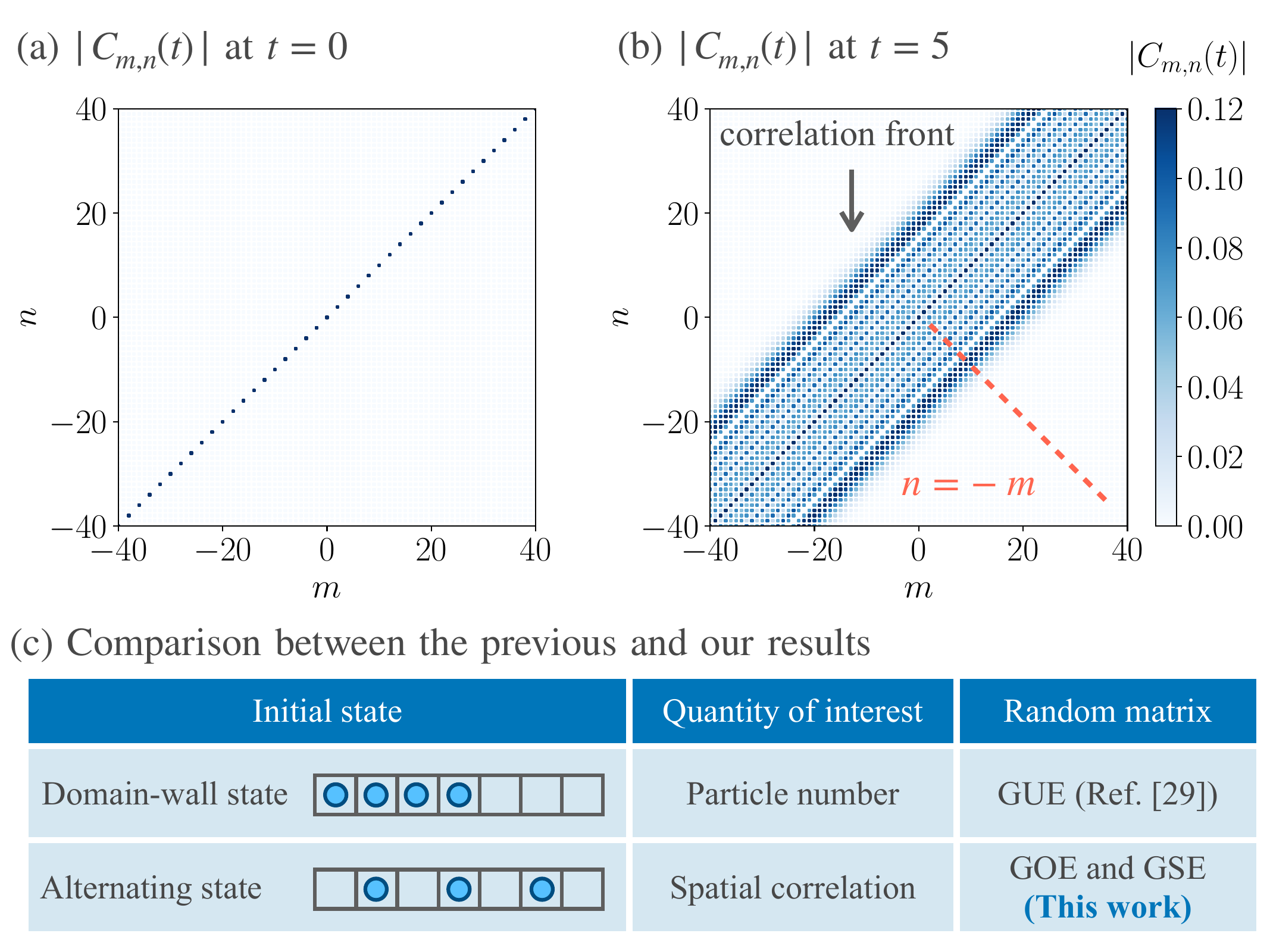}
\caption{
(a,b) Spatial distributions for modulus of a two-point correlator $C_{m,n}(t):= \bra{\psi(t)} \hat{a}_m^{\dagger} \hat{a}_n \ket{\psi(t)}$ with a quantum state $\ket{\psi(t)}$ at time (a) $t=0$ and (b) $t=5$. $\hat{a}_m^{\dagger}$ and $\hat{a}_m$ denote fermionic creation and annihilation operators at a site $m$. The initial state satisfies $C_{m,n}(0)= \delta_{m,n} (1 + (-1)^{n}) / 2$; no correlations exist for different sites. As time goes by, the correlator grows along a direction parallel to the dashed line corresponding to $n=-m$ in (b), and then the propagating correlation fronts emerge. (c) Table for the previous result~\cite{Eisler_2013} (second row) and ours (third row). The first, second, and third columns are initial states with the corresponding schematics, quantities of interest, and classes of random matrices associated with fermionic dynamics in Ref.~\cite{Eisler_2013} and this work. GUE, GOE, and GSE represent the Gaussian Unitary Ensemble, the Gaussian Orthogonal Ensemble, and the Gaussian Symplectic Ensemble, respectively \cite{mehta1,forrester2010}.
} 
\label{fig1} 
\end{center}
\end{figure}

In this Letter, we study the dynamical fluctuations around propagating correlation fronts in one-dimensional non-interacting fermions starting from an alternating initial state, where the particles occupy every other site. Figures~\ref{fig1} (a) and (b) display the time evolution of a two-point correlator $C_{m,n}(t)$ (precisely defined after Eq.~\eqref{initial}), from which one can clearly see the propagating correlation fronts. To explore universal aspects of the dynamical fluctuations, we introduce a cumulative correlation operator, which can capture the fluctuations around the fronts. Then, exactly solving the Schr\"odinger equation, we analytically show that, in the long-time dynamics, all the moments of the cumulative correlation operator are determined by universal correlation functions of the Gaussian Orthogonal Ensemble (GOE) and the Gaussian Symplectic Ensemble (GSE) at the soft edge in random matrix theory. Our main result is summarized in Table of Fig.~\ref{fig1}(c), where we emphasize differences between our findings and the previous results of Ref.~\cite{Eisler_2013}. The major contribution of this work to the research field of correlation dynamics is to uncover that the dynamical higher-order fluctuation around the correlation front can exhibit the universal behaviors featuring random matrix theory.

{\it Setup.\text{-}}
We consider non-interacting fermions on a one-dimensional lattice, and denote the fermionic annihilation and creation operators at a site $m \in \mathbb{Z}$ by $\hat{a}_m$ and $\hat{a}_m^{\dagger}$. Then, the Hamiltonian is defined by $ \hat{H} := -\sum_{m = -\infty}^{\infty} \left( \hat{a}_{m+1}^{\dagger} \hat{a}_m + \hat{a}_{m}^{\dagger} \hat{a}_{m+1}   \right)$. Under this setup, we can compute a quantum state $\ket{\psi(t)}$ at time $t$ by solving the Schr\"odinger equation ${\rm i} d \ket{\psi(t)}/dt = \hat{H} \ket{\psi(t)}$ with a given initial state $\ket{\psi(0)}$. Here we set $\hbar = 1$. The initial state used in this work is the alternating state $\ket{\psi_{\rm alt}}$, where the fermions occupy only the even sites (see the third row of Fig.~\ref{fig1}(c)):
\begin{align}
\ket{\psi_{\rm alt}} := \prod_{m=-\infty}^{\infty} \hat{a}_{2m}^{\dagger} \ket{0}
\label{initial}
\end{align}
with a vacuum $\ket{0}$. 

The quantity of our interest is the dynamical fluctuations around the propagating correlation fronts. Figures~\ref{fig1}(a) and (b) show the time evolution for the modulus of the two-point correlator $C_{m,n}(t):= \braket{ \hat{a}_m^{\dagger} \hat{a}_n }_{t}$ with $ \braket{\bullet}_t := \bra{\psi(t)} \bullet \ket{\psi(t)} $. One can see that the correlation fronts propagate along a direction parallel to the line $n=-m$ (see the dashed line in Fig.~\ref{fig1}(b)). To investigate the dynamical fluctuations around the fronts, we focus on the $2N$-point correlator on the line $n=-m$ defined by
\begin{eqnarray}
\left\langle\prod_{j=1}^{N} \hat{a}^{\dagger}_{m_j} \hat{a}_{-m_j}  \right\rangle_t
\label{multi_c}
\end{eqnarray}
with an integer $m_j ~(j \in \{1,\cdots, N \})$. This quantity captures the dynamical fluctuation around the correlation front when we appropriately choose $\{ m_1, \cdots, m_N  \}$ corresponding to the propagating front. As we will see shortly, the fluctuation of the correlation front is related to random matrix theory. To show this, it is convenient to consider a cumulative correlation operator $\hat{F}_{l}$ defined by
\begin{align}
\hat{F}_{l} := \sum_{m=l}^{\infty} \hat{a}^{\dagger}_m \hat{a}_{-m}, 
\label{cc_op}
\end{align}
where $l> 0$ is a positive integer. The moments of $\hat{F}_{l}$ capture the dynamical fluctuation around the front by appropriately choosing $l$. For the following calculations, we define a generating function for the moments by 
\begin{align}
Q(\lambda,t,l) := \braket{  e^{ \lambda \hat{F}_{l} } }_t
\label{G1}
\end{align}
with a real number $\lambda$. As shown in the following, the generating function includes the multipoint correlator~\eqref{multi_c}.

{\it Appearance of the GOE Tracy-Widom distribution.\text{-}}
Focusing on the propagating correlation front, we shall analytically show that $Q(\lambda,t,l)$ is related to the universal eigenvalue distribution function of random matrix theory, namely the GOE Tracy-Widom distribution \cite{GOETW,forrester2010}.

First, we derive a determinantal formula for $Q(\lambda,t,l)$. A straightforward calculation leads to
\begin{eqnarray}
Q(\lambda,t,l)  = 1 + \sum_{n=1}^{\infty} \sum_{ \substack{m_1 < m_2 < \cdots < m_n \\  m_{k} \in [ l,\infty)  } } \lambda^n \left\langle \prod_{j=1}^{n} \hat{a}^{\dagger}_{m_j} \hat{a}_{-m_j}  \right\rangle_t.
\label{Dform}
\end{eqnarray}
The Wick theorem enables us to decompose the $2n$-point correlator of Eq.~\eqref{Dform} into products of $C_{m,n}(t)$:
\begin{eqnarray}
\left\langle\prod_{j=1}^{n} \hat{a}^{\dagger}_{m_j} \hat{a}_{-m_j}  \right\rangle_t = {\rm det} \left[ C_{j,-k}(t) \right]_{j,k \in \{m_1,\cdots,m_n \}}. 
\label{Wick}
\end{eqnarray}
Substituting Eq.~\eqref{Wick} into Eq.~\eqref{Dform}, we obtain 
\begin{eqnarray}
Q(\lambda,t,l)  = {\rm det} \left[ \delta_{m,n} + \lambda C_{m,-n}(t) \right]_{ \ell^2[l,\infty) }
\label{Dform_f}
\end{eqnarray}
with the Kronecker delta $\delta_{m,n}$ and $\ell^2[l,\infty)$, the square-summable sequence space on $[l,\infty)$. The correlator $C_{m,n}(t)$ in Eq.~\eqref{Dform_f} is written as
\begin{eqnarray}
C_{m,n}(t) = \frac{1}{2} \delta_{m,n}  + \frac{1}{2} {\rm i}^{n+m}  J_{n-m}(4t) 
\label{C_con}
\end{eqnarray}
with the $n$-th order Bessel function of the first kind $J_{n}(x)$. The expression of the two-point correlator $C_{m,n}(t)$ in Eq.~\eqref{C_con} already appeared in Ref.~\cite{Flesch_2008} for non-interacting bosons. For completeness, we give its derivation in Sec.~I of Supplemental Material (SM) \cite{SM}. Combining Eqs.~\eqref{Dform_f} and~\eqref{C_con}, we obtain
\begin{eqnarray}
Q(\lambda,t,l)  =  {\rm det} \left[ \delta_{m,n} + \frac{\lambda}{2}  J_{n+m}(4t) \right]_{ \ell^2[l,\infty)}, 
\label{G1_det}
\end{eqnarray}
where we use $J_{-n}(x) = (-1)^n J_n(x)$ and then eliminate the trivial factor ${\rm i}^{m-n} (-1)^{n+m}$ by expanding the determinant.

Second, we derive the spatial form of the propagating correlation front by applying asymptotic analysis to Eq.~\eqref{C_con} for studying the dynamical fluctuation around the front. We here focus on the front along the line $n=-m < 0$, namely $C_{m,-m}(t)$. Introducing a variable $x$ by  $m= \lfloor 2 t + x (2 t)^{1/3} /2  \rfloor$ with the floor function $ \lfloor \bullet \rfloor $, we can show, for $t \gg 1$, 
\begin{eqnarray}
C_{m,-m}(t) \simeq \frac{1}{2(2t)^{1/3}} {\rm Ai}(x), 
\label{cfront}
\end{eqnarray}
where we use the asymptotic formula $J_{ \lfloor 4 t + (2 t)^{1/3} x  \rfloor }(4t) \simeq {\rm Ai}(x)/(2t)^{1/3}~(t \gg 1)$ with the Airy function ${\rm Ai}(x)$ \cite{abramowitz1988handbook} (see Sec. II of \cite{SM} for the derivation). This shows that the peak of $C_{m,-m}(t)~(m > 0)$ is approximately given by $(m,n) \simeq (2t, -2t)$, consistent with the fact that the front ballistically propagates with the maximal group velocity $\underset{k}{{\rm max}} \{ d \epsilon(k)/dk \} = 2$, as shown in Fig.~\ref{fig1}(b). Here, $\epsilon(k) = -2 \cos(k)$ is the energy eigenvalue of our model.

Finally, we study the generating function $Q(\lambda,t,l)$ around the propagating correlation front. Taking into account the fact that the front moves with the velocity $2$, we choose $l = l_{t,s} := \lfloor 2t + s(2t)^{1/3}/2 \rfloor$ with a rescaled coordinate $s$ and introduce two rescaled coordinates $x$ and $y$ by $n= \lfloor 2 t + x (2 t)^{1/3} /2  \rfloor$ and $m= \lfloor 2 t + y (2 t)^{1/3} /2  \rfloor$ in Eq.~\eqref{G1_det}. Then, the moments of $\hat{F}_{l_{t,s}}$ characterize the fluctuation around the front. Under this setup, we take $\lambda = -2$ and use the asymptotic formula employed in the derivation of Eq.~\eqref{cfront}, getting 
\begin{eqnarray}
Q(-2,t, l_{t,s})  \simeq {\rm det} \left[ 1 - \frac{1}{2}{\rm Ai} \left( \frac{x+y}{2} \right) \right]_{ {L}^2(s,\infty)}
\label{G1_det_scaling}
\end{eqnarray}
for $t \gg 1$. This form is identical to a determinantal formula for the GOE Tracy-Widom distribution \cite{Sasamoto_2005,Ferrari_2005}, which is a universal cumulative distribution function for the largest eigenvalue for GOE. 

Our result of Eq.~\eqref{G1_det_scaling} strongly suggests that the dynamical fluctuation around the propagating correlation front is related to universal eigenvalue distributions of random matrix theory. In the rest of the paper, we further explore the detailed connection of the fluctuation to random matrix theory, by focusing on the $n$-th moment $M_{n}(t, l_{t,s}) := \left\langle  \left( \hat{F}_{l_{t,s}} \right)^{n}  \right\rangle_t$.

{\it Connection of random matrix theory to the moments.\text{-}}
Employing the analytical method used in the derivation of Eq.~\eqref{G1_det_scaling}, we shall show that universal correlation functions of GOE and GSE asymptotically determine all the $n$-th moments $M_{n}(t,l_{t,s})$.

We first introduce notations of random matrix theory before the detailed analysis. Let us denote $n$-point eigenvalue correlation functions for GOE and GSE by $R_{n}^{\rm GOE}(x_1,\cdots, x_n)$ and $R_{n}^{\rm GSE}(x_1,\cdots, x_n)$. We suppose that $G_{\rm GOE}(\lambda,s)$ and $G_{\rm GSE}(\lambda,s)$ represents the generating functions of gap probabilities in the region $(s,\infty)$. By definition, they satisfy
\begin{eqnarray}
\left. \frac{d^n G_{\alpha}(\lambda,s) }{d\lambda^n} \right|_{\lambda=0}  = (-1)^n \int_{s}^{\infty} dx_1 \cdots \int_{s}^{\infty} dx_n R_n^{\alpha}(x_1, \cdots, x_n) \nonumber \\
\label{GSOE_diff}
\end{eqnarray}
with $\alpha$ taking GOE or GSE \cite{mehta1,forrester2010}. We summarize their detailed definitions and properties in Sec.~III of SM~\cite{SM}. In Ref.~\cite{Bornemann_2010}, Bornemann showed that, under the soft-edge scaling limit, the generating functions have determinantal formulas,
\begin{eqnarray}
&& G_{\rm GSE}(\lambda,s) =  \frac{1}{2} H \left( \sqrt{\lambda},s \right) + \frac{1}{2} H \left( -\sqrt{\lambda},s \right),  \\
&&G_{\rm GOE}(\lambda,s) = \frac{1}{2} \left( 1 + \sqrt{ \frac{\lambda}{2-\lambda} }  \right) H \left( \sqrt{\lambda(2-\lambda)},s \right) \nonumber  \\
&& \hspace{15mm} + \frac{1}{2} \left( 1 - \sqrt{ \frac{\lambda}{2-\lambda} }  \right) H \left( - \sqrt{\lambda(2-\lambda)},s \right) 
\end{eqnarray}
with $H(z,s) := {\rm det} \left[  1 - \frac{z}{2}{\rm Ai} \left( \frac{x+y}{2} \right) \right]_{ {L}^2(s,\infty)}$ and real variables $\lambda$ and $s$. 

To see the connection between our correlation-front dynamics and random matrix theory, let us define two functions for the 
cumulative correlation operator $\hat{F}_l$ as
\begin{eqnarray}
&&G_1(\lambda, t,l) :=  \left\langle \cosh  \left( 2 \sqrt{ \lambda } \hat{F}_{l}  \right)  \right\rangle_t,  \label{G2} \\
&&G_2(\lambda, t,l) :=  \left\langle \cosh  \left( 2 \sqrt{ \lambda(2-\lambda)} \hat{F}_{l}  \right)  \right\rangle_t \nonumber \\
&& ~~~~~~~~~~~~~~ - \sqrt{\frac{\lambda}{2-\lambda}}  \left\langle \sinh  \left( 2 \sqrt{ \lambda(2-\lambda)} \hat{F}_{l}  \right)  \right\rangle_t.~~~~~~ \label{G3}
\end{eqnarray}
Expanding them with $\lambda$, we find that $G_1(\lambda, t,l)$ consists of the even-order moments $M_{2n}(t,l)$ while $G_2(\lambda, t,l)$ does of all the moments $M_{2n}(t,l)$ and $M_{2n+1}(t,l)$. Note that $G_1(\lambda, t,l)$ and  $G_2(\lambda, t, l)$ are the power series with $\lambda$ and thus we can differentiate them at $\lambda=0$. Using the same asymptotic analysis used in the derivation of Eq.~\eqref{G1_det_scaling}, we can show, for $t \gg 1$,
\begin{eqnarray}
&&G_1(\lambda, t, l_{t,s}) \simeq  G_{\rm GSE}(\lambda,s),  \label{G2_scaling} \\
&&G_2(\lambda, t,l_{t,s}) \simeq   G_{\rm GOE}(\lambda,s)  \label{G3_scaling}.
\end{eqnarray}
These are the fundamental relations, establishing that the dynamical fluctuation of the propagating correlation front in the long-time regime is described by the universal correlation functions of GOE and GSE at the soft edge, as shown in the subsequent paragraphs. The detailed derivation of Eqs.~(\ref{G2_scaling}) and (\ref{G3_scaling}) is given in Secs.~IV and V of SM \cite{SM}.

We next differentiate $G_1(\lambda, t, l_{t,s})$ with respect to $\lambda$, getting
\begin{eqnarray}
\left. \frac{d^n G_1(\lambda, t, l_{t,s})}{d\lambda^{n}} \right| _{\lambda=0} = \frac{4^n n!}{(2n)!} M_{2n}(t, l_{t,s})
\label{G2_dif}
\end{eqnarray}
as derived in Sec.~VI of SM \cite{SM}. Using Eqs.~\eqref{GSOE_diff}, \eqref{G2_scaling}, and \eqref{G2_dif}, we obtain
\begin{eqnarray}
&& M_{2n}(t, l_{t,s}) \nonumber \\
&&  \simeq \frac{(-1)^n (2n)!}{4^n n!}\int_{s}^{\infty} dx_1 \cdots \int_{s}^{\infty} dx_n R_n^{\rm GSE}(x_1, \cdots, x_n) \nonumber \\
\label{evenmoments}
\end{eqnarray}
for $t \gg 1$. Thus we elucidate that all the even-order moments $M_{2n}(t,l_{t,s})$ are asymptotically determined by the universal correlation function $R_n^{\rm GSE}(x_1, \cdots, x_n)$ for GSE.

Following the same procedure just above, we differentiate $G_2(\lambda, t, l_{t,s}) $ with respect to $\lambda$ and then obtain
\begin{eqnarray}
&~&  \sum_{k=0}^{\lfloor \frac{n}{2} \rfloor} \frac{  (-1)^{k + n } 2^{3n-4k} (n-k)!  n!   }{(2n-2k)! (n-2k)! k!}  M_{2n-2k}(t, l_{t,s})  \nonumber \\
&-& \sum_{k=0}^{\lfloor \frac{n-1}{2} \rfloor} \frac{  (-1)^{k + n}2^{3n-4k-2} (n-k-1)!  n!  }{(2n-2k-1)! (n-2k-1)! k!}  M_{2n-2k-1}(t, l_{t,s})  \nonumber \\
&\simeq& \int_{s}^{\infty} dx_1 \cdots \int_{s}^{\infty} dx_n R_n^{\rm GOE}(x_1, \cdots, x_n)
\label{oddmoments}
\end{eqnarray}
for $t \gg 1$ (see Sec.~VII of SM \cite{SM} for the detailed derivation). Using Eqs.~\eqref{evenmoments} and \eqref{oddmoments}, we can recursively demonstrate that all the odd-order moments $M_{2n+1}(t,l_{t,s})$ are expressed by the universal correlation functions $R_n^{\rm GOE}(x_1, \cdots, x_n)$ and $R_n^{\rm GSE}(x_1, \cdots, x_n)$ for GOE and GSE. For example, putting $n=1$ into Eqs.~\eqref{evenmoments} and \eqref{oddmoments}, we obtain
\begin{eqnarray}
&&M_{1}(t, l_{t,s}) \simeq \cfrac{1}{2} \int_{s}^{\infty} dx \left(  R_1^{\rm GOE}(x) - 2 R_1^{\rm GSE}(x)  \right), 
\label{firstmoment2} \\
&&M_2(t, l_{t,s}) \simeq -\frac{1}{2} \int_{s}^{\infty} dx  R_1^{\rm GSE}(x).
\label{secondmoment1}
\end{eqnarray}

\begin{figure}[t]
\begin{center}
\includegraphics[keepaspectratio, width=8.7cm]{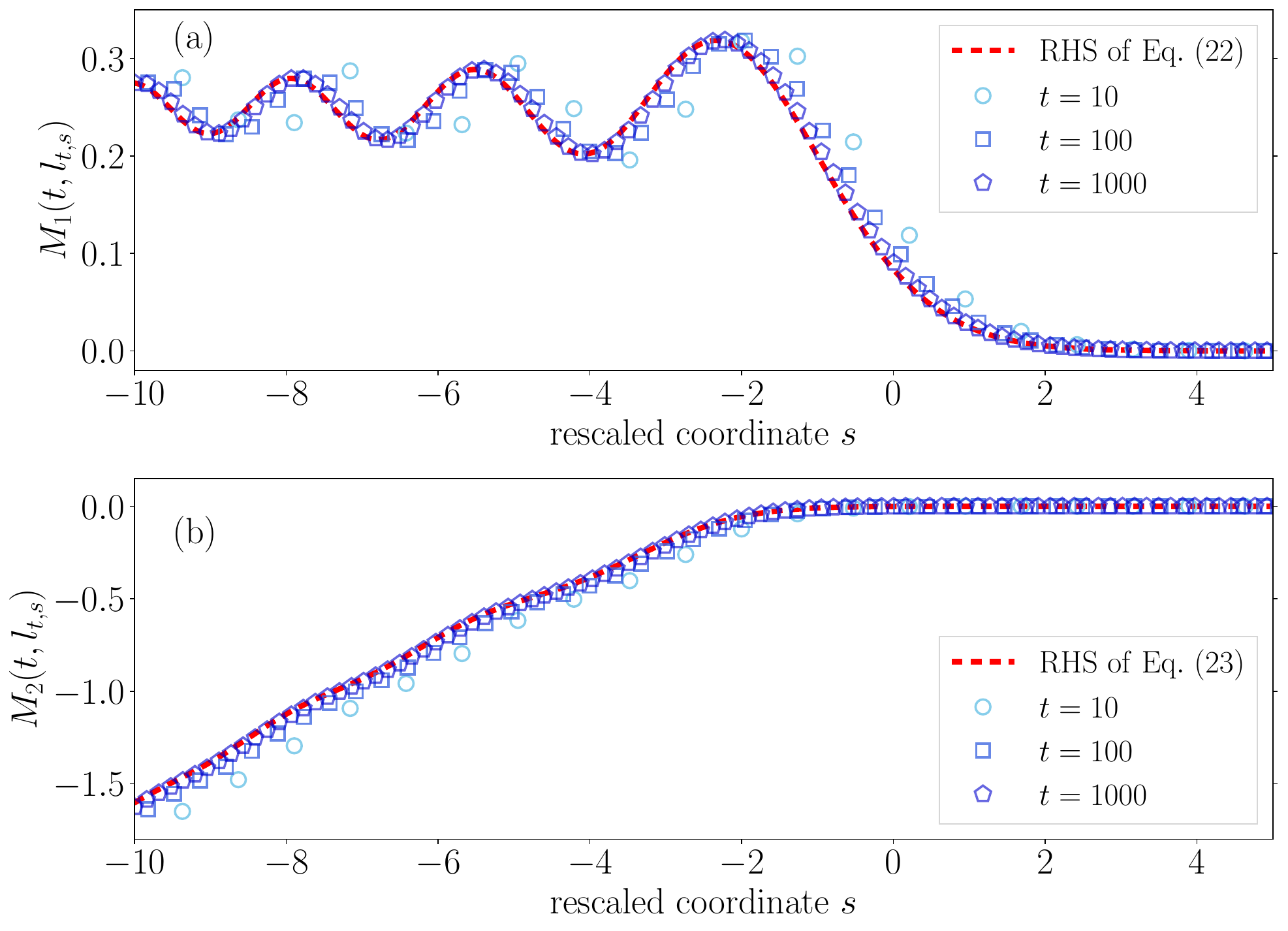}
\caption{
Numerical vertification of Eqs.~\eqref{firstmoment2} and \eqref{secondmoment1}. The circle, square, and pentagon markers denote (a) $M_1(t, l_{t,s})$ and (b) $M_2(t, l_{t,s})$ at $t=10,100$, and $1000$, respectively. The rescaled coordinate $s$ is defined through $l_{t,s} = \lfloor 2t + s(2t)^{1/3}/2 \rfloor$. The dashed lines in (a) and (b) represent the right-hand sides (RHSs) of Eqs.~\eqref{firstmoment2} and \eqref{secondmoment1}, respectively.
} 
\label{fig2} 
\end{center}
\end{figure}

We numerically verify Eqs.~\eqref{firstmoment2} and \eqref{secondmoment1} by solving the Schr\"odinger equation. Figure~\ref{fig2} displays the time evolution of $M_1(t, l_{t,s})$ and $M_2(t, l_{t,s})$. We find that Eqs.~\eqref{firstmoment2} and \eqref{secondmoment1} hold well for $t \gg 1$.

{\it Discussion.\text{-}} 
We discuss (i) dependence on initial states, (ii) experimental possibility, and (iii) interaction effect for our results.

Let us first consider the topic (i). As described in Sec.~VIII of SM~\cite{SM}, we analytically and numerically investigate the dependence on initial states, showing that GOE and GSE can characterize all the moments under appropriate rescaling even when using a modified alternating state with a filling factor different from $1/2$. We also find several exceptional initial states, spatial configurations of which are similar to the alternating state but the GOE and GSE behaviors do not appear. We identify a condition for this absence of the GOE and GSE behaviors, which is given by Eq.~(S-69) in SM~\cite{SM}. We find that the number of such exceptional states is smaller than that of the initial states exhibiting the GOE and GSE behaviors (see the details of Sec.~VIII C of SM~\cite{SM}).

As to the topic (ii), we need to observe the multipoint correlator~\eqref{multi_c} to verify our theoretical prediction, but its observation is generally difficult. To our knowledge, in Ref.~\cite{takasu2020}, Takasu et~al. experimentally obtained the two-point correlator of bosonic dynamics in optical lattices by observing the momentum distribution via time-of-flight images, while there are no such works for fermions. In the future, it may be possible to observe the fermionic correlator in cold atom experiments, which will enable us to explore our theoretical prediction.

We discuss the topic (iii). In the previous work~\cite{Collura_2018}, Collura et al. theoretically studied the interaction effect on the result of Ref.~\cite{Eisler_2013} where Eisler and R\'acz reported that GUE characterizes the propagating particle-density front in non-interacting fermions. Using the density-matrix-renormalization-group method, they numerically found that the GUE behavior disappeared in the interacting system. Thus, it is interesting to discuss our results in interacting systems. We, however, will leave it as a future work since it will be pretty demanding to numerically access the long-time dynamics starting from the alternating state.  

{\it Conclusions and prospects.\text{-}} 
We theoretically considered the non-interacting fermions on the one-dimensional lattice, studying the dynamics starting from the alternating initial state, where the particles are on every other site. In this case, the two-point correlator $C_{m,n}(t)$ formed the propagating correlation front, as shown in Fig.~\ref{fig1}(b). Focusing on the dynamical fluctuation around the front in the late stage of the dynamics, we analytically showed that the universal correlation functions of GOE and GSE asymptotically determine all the moments $M_n(t, l_{t,s})$ for the cumulative correlation operators capturing the fluctuation around the propagating correlation front. We further studied the dependence of our results on initial states, finding that the GOE and GSE behaviors can survive for the modified alternating states. Thus our result is universal in that the behaviors emerge when systems are mapped into the non-interacting fermions with initial states similar to the alternating state.

As a prospect, it will be intriguing to study dynamical fluctuations around propagating fronts for other physical quantities. Thanks to the light-cone propagation, propagating fronts often emerge even for spins, particle numbers, and local energy. Connections of random matrix theory to such unexplored dynamical fluctuations around the fronts are of great interest. 

Another interesting direction is to explore relations between classical stochastic processes and our results. Universal distributions of random matrix theory have been intensively investigated in classical stochastic processes, examples of which include the Kardar-Parisi-Zhang equation, asymmetric simple exclusion processes, and polynuclear growth models \cite{PS2002,Sasamoto2007,Krug2010,CORWIN2012,Quastel2015,sasamoto2016,takeuchi2018}. For example, in a totally asymmetric exclusion process starting from the alternating state, the particle transport features the GOE Tracy-Widom distribution. This classical dynamics may be related to our result. Thus, it is fundamentally interesting to pursue relations between classical stochastic processes and correlation dynamics in quantum systems from the unified perspective of random matrix theory.

\begin{acknowledgments}
The authors are grateful to Masaya Kunimi and Hiroki Moriya for the fruitful discussion, and Ryusuke Hamazaki for the helpful comments on the manuscript. 
The work of KF has been supported by JSPS KAKENHI Grant No. JP23K13029. 
The work of TS has been supported by JSPS KAKENHI Grants No. JP21H04432, No. JP22H01143.
\end{acknowledgments}
\bibliography{reference}

\widetext
\clearpage

\setcounter{equation}{0}
\setcounter{figure}{0}
\setcounter{section}{0}
\setcounter{table}{0}
\renewcommand{\theequation}{S-\arabic{equation}}
\renewcommand{\thefigure}{S-\arabic{figure}}
\renewcommand{\thetable}{S-\arabic{table}}

\section*{Supplemental Material for ``Random Matrix Statistics in Propagating Correlation Fronts of Fermions''}

\centerline{Kazuya Fujimoto and Tomohiro Sasamoto}
\vspace{3mm}

\centerline{Department of Physics, Tokyo Institute of Technology, 2-12-1 Ookayama, Meguro-ku, Tokyo 152-8551, Japan}
\vspace{5mm}

\par\vskip3mm \hrule\vskip.5mm\hrule \vskip.30cm
This Supplemental Material describes the following:
\begin{itemize}
\item[  ]{ (I) Derivation of Eq.~\eqref{C_con}, } 
\item[  ]{ (II) Asymptotic analysis for the Bessel function of the first kind, } 
\item[  ]{ (III) Generating functions in random matrix theory, } 
\item[  ]{ (IV) Derivation of Eq.~\eqref{G2_scaling}, } 
\item[  ]{ (V) Derivation of Eq.~\eqref{G3_scaling}, } 
\item[  ]{ (VI) Derivation of Eq.~\eqref{G2_dif}, } 
\item[  ]{ (VII) Derivation of Eq.~\eqref{oddmoments}, } 
\item[  ]{ (VIII) Dependence on initial states. } 
\end{itemize}
\par\vskip1mm \hrule\vskip.5mm\hrule

\vspace{5mm}

\section{Derivation of Eq.~(8)}
We explain how to derive Eq.~\eqref{C_con} in the main text. Suppose that we have non-interacting fermions on a finite one-dimensional lattice labeled by $\Lambda=\{-L,-L+1, \cdots, L \}$ with a positive even integer $L$. The fermionic annihilation and creation operators at a site $m \in \Lambda$ are denoted by $\hat{a}_m$ and $\hat{a}_m^{\dagger}$. Then, we define the Hamiltonian as 
\begin{align}
\hat{H} := -\sum_{m = -L}^{L-1} \left( \hat{a}_{m+1}^{\dagger} \hat{a}_m + \hat{a}_{m}^{\dagger} \hat{a}_{m+1}   \right), 
\label{S_H}
\end{align}
where the boundary condition is assumed to be periodic ($\hat{a}_{L}=\hat{a}_{-L}$). The initial state $\ket{\psi(0)} $ is the alternating state given by
\begin{align}
\ket{\psi(0)} := \prod_{m=-L/2}^{L/2-1} \hat{a}_{2m}^{\dagger} \ket{0}
\label{S_initial}
\end{align}
with a vacuum $\ket{0}$.  Under this setup, the quantum state $\ket{\psi(t)}$ at time $t$ obeys the Schr\"odinger equation ${\rm i} d  \ket{\psi(t)} /dt = \hat{H} \ket{\psi(t)}$ where we set $\hbar = 1$.  

We solve the Heisenberg equation for the annihilation operator, obtaining an explicit form of the two-point correlator $C_{m,n}(t) := \braket{\hat{a}_m^{\dagger} \hat{a}_n }_t$ in the thermodynamic limit ($L \rightarrow \infty$). Here, we introduce the notation $\braket{\bullet}_t := \bra{\psi(t)} \bullet \ket{\psi(t)}$. Applying the discrete Fourier transformation to the annihilation operator $\hat{a}_{m}(t) := e^{ {\rm i}\hat{H} t } \hat{a}_m e^{ -{\rm i}\hat{H} t }$ in the Heisenberg picture, we can obtain 
\begin{align}
\hat{a}_{m}(t) = \frac{1}{\sqrt{2L}} \sum_{\alpha=-L}^{L-1} \hat{A}_{\alpha} \exp \biggl( {\rm i} 2t \cos( \delta k \alpha ) + {\rm i} \delta k \alpha m \biggl)
\label{S_af}
\end{align}
with the wavenumber unit $\delta k := \pi/L$ and the initial annihilation operator $\hat{A}_{\alpha}$ in the wavenumber space, defined as
\begin{align}
\hat{A}_{\alpha} := \frac{1}{\sqrt{2L}}\sum_{m=-L}^{L-1} \hat{a}_m \exp \biggl( -{\rm i} \delta k \alpha m \biggl).
\end{align}
Substituting Eq.~\eqref{S_af}  into the definition of $C_{m,n}(t)$, we get 
\begin{align}
C_{m,n}(t) = \frac{1}{2L}  \sum_{\alpha=-L}^{L-1} \sum_{\beta=-L}^{L-1} \bra{\psi(0)} {\hat{A}_{\alpha}^{\dagger} \hat{A}_{\beta}} \ket{\psi(0)} \exp \biggl(  {\rm i} 2t \bigl( \cos( \delta k \beta ) - \cos( \delta k \alpha ) \bigl)  + {\rm i} \delta k \bigl(\beta n - \alpha m \bigl) \biggl).
\label{S_C_dis0}
\end{align}
The quantum average of the creation and annihilation operators in the wavenumber space becomes
\begin{align}
\bra{\psi(0)} \hat{A}_{\alpha}^{\dagger} \hat{A}_{\beta} \ket{\psi(0)} = \frac{1}{2} \delta_{\alpha, \beta-L} + \frac{1}{2} \delta_{\alpha, \beta} +  \frac{1}{2} \delta_{\alpha, \beta+L}, 
\label{S_A}
\end{align}
which can be derived using the Eq.~\eqref{S_initial} of the alternating initial state. We use Eqs.~\eqref{S_C_dis0} and \eqref{S_A}, obtaining
\begin{align}
C_{m,n}(t) = \frac{1}{2} \delta_{m,n} + \frac{(-1)^n}{4L} \sum_{\alpha=-L}^{L-1}  \exp \left[ -{\rm i} 4 t \cos( \delta k \alpha ) + {\rm i} \delta k \alpha (n-m) \right].
\label{S_C_dis}
\end{align}
Taking the thermodynamic limit ($L \rightarrow \infty$) for Eq.~\eqref{S_C_dis}, we derive
\begin{eqnarray}
C_{m,n}(t) = \frac{1}{2} \delta_{m,n} + \frac{1}{2} {\rm i}^{n+m} J_{n-m}(4t) 
\label{S_C_con}
\end{eqnarray}
with the $n$-th order Bessel function of the first kind $J_{n}(x)$ \cite{Flesch_2008}. In this derivation, we use the following integral formula given by
\begin{align}
J_{n}(x) = \frac{1}{ 2 \pi {\rm i}^n }\int_{0}^{2 \pi} d \theta \exp \biggl( {\rm i} x \cos(\theta) + {\rm i} n \theta \biggl). 
\label{S_bessel}
\end{align}
Note that an origin of the Bessel function of the first kind is the contribution coming from $\alpha = \beta \pm L $ in Eq.~\eqref{S_A}. This fact becomes important when discussing dependence of our results on initial states as described in Sec.~\ref{DIS}. 

\section{Asymptotic analysis for the Bessel function of the first kind}
We shall show that the Bessel function of the first kind $J_{n}(4t)$ with $n= \lfloor 4 t + (2 t)^{1/3} x  \rfloor $ asymptotically approaches the Airy function for $t \gg 1$ \cite{abramowitz1988handbook}. Here, $\lfloor \bullet \rfloor$ denotes the floor function. The integral representation of Eq.~\eqref{S_bessel} leads to 
\begin{align}
J_{ \lfloor 4 t + (2 t)^{1/3} x  \rfloor }(4 t) \simeq \frac{1}{ 2 \pi {\rm i}^{ 4 t + (2 t)^{1/3} x  } }\int_{0}^{2 \pi} d \theta \exp \biggl( {\rm i} 4t \bigl(\cos\theta + \theta \bigl)  + {\rm i} (2t)^{1/3} x \theta \biggl), 
\label{S_Asym1}
\end{align}
where we neglect the difference $\lfloor 4 t + (2 t)^{1/3} x  \rfloor -4 t - (2 t)^{1/3} x$ since it becomes very small compared with $ 4 t + (2 t)^{1/3} x $ for $t \gg 1$. The integrand in Eq.~\eqref{S_Asym1} can rapidly oscillate for $t \gg 1$ since the term with $f(\theta) := \cos \theta + \theta$ is proportional to time $t$. Then, employing the conventional stationary phase method, we investigate $f(\theta)$ and then find that it has an extremum at $\theta_0 := \pi/2$. Thus, the contribution around $\theta_0$ to the integral of Eq.~\eqref{S_Asym1} becomes dominant for $t \gg 1$. 
Using the expansion $f(\theta) = \pi/2 + (\theta - \theta_0)^3/6 + \mathcal{O}( (\theta - \theta_0)^{5})$, we can approximate the Bessel function of the first kind as 
\begin{align}
J_{ \lfloor 4 t + (2 t)^{1/3} x  \rfloor }(4 t) \simeq \frac{1}{ \pi  } \int_{0}^{\infty} d \theta \cos \left(  (2t)^{1/3} x \theta + \frac{2t}{3} \theta^3  \right).
\label{S_Asym2}
\end{align}
Then, this gives 
\begin{align}
J_{ \lfloor 4 t + (2 t)^{1/3} x  \rfloor }(4 t) \simeq \frac{1}{  (2t)^{1/3} } {\rm Ai}(x), 
\label{S_Asym3}
\end{align}
where we use the following formula for the Airy function \cite{airy_func}:
\begin{align}
{\rm Ai}(x) = \frac{1}{\pi} \int_{0}^{\infty}  d \theta \cos \left( x \theta + \frac{ \theta^3}{3} \right).
\label{S_Asym4}
\end{align}

\section{Generating functions in random matrix theory}
We briefly review generating functions in random matrix theory \cite{mehta1,forrester2010}. The definitions of random matrices used in our work are based on Ref.~\cite{Bornemann_2010}.

Suppose that we have $M \times M$ random matrices sampled by the Gaussian Orthogonal Ensemble (GOE) or the Gaussian Symplectic Ensemble (GSE), and denote probability density functions for the eigenvalues $(x_1, x_2, \cdots, x_M)$ by $P^{\rm GOE}(x_1,x_2,\cdots, x_M)$ and $P^{\rm GSE}(x_1,x_2,\cdots, x_M)$, respectively. Then, the generating function with a real parameter $\lambda$ is defined by
\begin{eqnarray}
G_{\alpha}(\lambda,s) := \int^{\infty}_{-\infty} dx_1 \cdots \int^{\infty}_{-\infty} dx_M  P^{\alpha}(x_1,x_2,\cdots, x_M) \prod_{n=1}^{M} \biggl( 1 - \lambda \chi_D(x_n) \biggl), 
\label{S_8}
\end{eqnarray}
where $\chi_D(x)$ is an indication function with $D := [s,\infty)$ and a real number $s$. The label $\alpha$ takes GOE or GSE hereafter in this section. 

We next introduce an $n$-point eigenvalue correlation function and an $n$-point gap probability denoted by $R_{n}^{\alpha}(x_1,x_2,\cdots, x_n)$ and $E_n^{\alpha} (s,\infty)$, respectively.
These quantities are defined as
\begin{eqnarray}
R_{n}^{\alpha}(x_1,x_2,\cdots, x_n) &:=& \frac{M!}{(M-n)!} \int_{-\infty}^{\infty} \left( \prod_{k=n+1}^M dx_k \right)  P^{\alpha}(x_1,x_2,\cdots, x_M),     \\
E_n^{\alpha} (s,\infty) &:=& \frac{M!}{(M-n)! n!} \int^{\infty}_s dx_1 \cdots \int_{s}^{\infty} dx_n \int_{-\infty}^s dx_{n+1} \cdots \int_{-\infty}^s dx_{M} P^{\alpha}(x_1,x_2,\cdots, x_M) . 
\end{eqnarray}
Then, the generating function of Eq.~\eqref{S_8} satisfies
\begin{eqnarray}
G_{\alpha}(\lambda,s) &=& 1 + \sum_{n=1}^{M} \frac{(-\lambda)^n}{n!} \int_{s}^{\infty} dx_1 \int_{s}^{\infty} dx_2 \cdots \int_{s}^{\infty} dx_n R^{\alpha}_{n}(x_1,x_2,\cdots, x_n) \label{S_9} \\
&=& \sum_{n=0}^{M} (1-\lambda)^n E_{n}^{\alpha}(s,\infty).
\end{eqnarray}
Differentiating Eq.~\eqref{S_9} with respect to $\lambda$, we obtain 
\begin{eqnarray}
\left. \frac{d^n}{d\lambda^n} G_{\alpha}(\lambda,s) \right|_{\lambda=0} = (-1)^n \int_{s}^{\infty} dx_1 \cdots \int_{s}^{\infty} dx_n R_n^{\alpha}(x_1, \cdots, x_n), 
\label{S_GSOE_diff}
\end{eqnarray}
which is Eq.~\eqref{GSOE_diff} in the main text. 

Explicit expressions of $R_1^{\rm GOE}(x) $ and $R_1^{\rm GSE}(x) $ under the soft-edge scaling limit are given by 
\begin{eqnarray}
&& R_1^{\rm GSE}(x) = \cfrac{1}{2} \left( \cfrac{d}{dx}{\rm Ai}(x) \right)^2 - \cfrac{{\rm Ai}(x)}{2}   \cfrac{d^2}{dx^2} {\rm Ai}(x)  - \cfrac{{\rm Ai}(x)}{4}  \int_x^{\infty} {\rm Ai}(y) dy, \\
&& R_1^{\rm GOE}(x) = \left( \cfrac{d}{dx}{\rm Ai}(x) \right)^2 - {\rm Ai}(x)  \cfrac{d^2}{dx^2} {\rm Ai}(x)  - \cfrac{{\rm Ai}(x)}{2}  \int_x^{\infty} {\rm Ai}(y) dy + \cfrac{1}{2} {\rm Ai}(x).
\end{eqnarray}
We use these expressions in Fig.~\ref{fig2} of the main text.

\section{Derivation of Eq.~(17) }
We derive Eq.~\eqref{G2_scaling} of the main text by employing the analytical technique used in deriving the GOE Tracy-Widom distribution in the correlation dynamics. We can rewrite Eq.~\eqref{G2} of the main text as
\begin{eqnarray}
G_1(\lambda, t, l) &=&  \left\langle \cosh  \left( 2 \sqrt{\lambda} \hat{F}_{l}  \right)  \right\rangle_t \\
&=& \frac{1}{2} \left\langle e^{  2 \sqrt{\lambda} \hat{F}_{l}  }  \right\rangle_t  + \frac{1}{2}   \left\langle e^{ -2 \sqrt{\lambda} \hat{F}_{l}  }  \right\rangle_t  \\
&=& \frac{1}{2} Q \left( 2\sqrt{\lambda},t,l \right) + \frac{1}{2} Q \left(-2\sqrt{\lambda},t,l \right), 
\label{S1}
\end{eqnarray}
where we introduce the notation $Q(\lambda,t,l) := \braket{  e^{ \lambda \hat{F}_{l} } }_t$. From Eq.~\eqref{G1_det} of the main text, the expression of $Q \left( \pm 2 \sqrt{\lambda},t,l \right)$ becomes
\begin{eqnarray}
Q \left( \pm 2 \sqrt{\lambda},t, l \right) = {\rm det} \left[ \delta_{m,n} \pm \sqrt{\lambda} J_{n+m} (4t) \right]_{ \ell^2[l,\infty) }.
\end{eqnarray}
Introducing a rescaled coordinate $s$ as $l = l_{t,s} := \lfloor 2t + s (2t)^{1/3}/2 \rfloor$ used in the main text, we get 
\begin{eqnarray}
Q \left( \pm 2 \sqrt{\lambda},t, l_{t,s} \right) \simeq {\rm det} \left[ 1 \pm \frac{\sqrt{\lambda}}{2} {\rm Ai}\left( \frac{x+y}{2} \right) \right]_{L^2(s,\infty)}~~~~(t \gg 1).
\label{S2}
\end{eqnarray}
Substituting Eq.~\eqref{S2} into Eq.~\eqref{S1}, we obtain
\begin{eqnarray}
G_1 \left( \lambda, t, l_{t,s} \right) \simeq \frac{1}{2} {\rm det} \left[ 1 + \frac{\sqrt{\lambda}}{2} {\rm Ai}\left( \frac{x+y}{2} \right) \right]_{L^2(s,\infty)} + \frac{1}{2} {\rm det} \left[ 1 - \frac{\sqrt{\lambda}}{2} {\rm Ai}\left( \frac{x+y}{2} \right) \right]_{L^2(s,\infty)}
\label{S3}
\end{eqnarray}
for $t \gg 1$. The right-hand side is identical to the generating function for  GSE (see Eq.~(6.10) in Ref.~\cite{Bornemann_2010}).

\section{Derivation of Eq.~(18) }
We derive Eq.~\eqref{G3_scaling} of the main text by following the almost same procedure in the derivation of Eq.~\eqref{S3}. 
Equation~\eqref{G3} of the main text can be written as
\begin{eqnarray}
G_2(\lambda, t,l) &=&  \left\langle \cosh  \left( 2 \sqrt{\lambda(2-\lambda)} \hat{F}_{l}  \right)  \right\rangle_t   - \sqrt{\frac{\lambda}{2-\lambda}}  \left\langle \sinh  \left( 2 \sqrt{\lambda(2-\lambda)} \hat{F}_{l}  \right)  \right\rangle_t \nonumber \\
&=&  \frac{1}{2}  \left\langle \exp  \left( 2 \sqrt{\lambda(2-\lambda)} \hat{F}_{l}  \right)  \right\rangle_t  + \frac{1}{2}   \left\langle \exp  \left( -2 \sqrt{\lambda(2-\lambda)} \hat{F}_{l}  \right)  \right\rangle_t \nonumber \\
&-& \frac{1}{2}\sqrt{\frac{\lambda}{2-\lambda}}   \left\langle \exp  \left( 2 \sqrt{\lambda(2-\lambda)} \hat{F}_{l}  \right)  \right\rangle_t  + \frac{1}{2} \sqrt{\frac{\lambda}{2-\lambda}} \left\langle \exp  \left( -2 \sqrt{\lambda(2-\lambda)} \hat{F}_{l}  \right)  \right\rangle_t  \\
&=& \frac{1}{2}\left( 1 + \sqrt{\frac{\lambda}{2-\lambda}} \right)  Q(-2 \sqrt{\lambda(2-\lambda)},t,l) + \frac{1}{2}\left( 1 - \sqrt{\frac{\lambda}{2-\lambda}}\right) Q(2\sqrt{\lambda(2-\lambda)},t,l).
\label{S4}
\end{eqnarray}
Using Eq.~\eqref{S2}, we obtain 
\begin{eqnarray}
G_2(\lambda, t, l_{t,s}) &\simeq&  \frac{1}{2}\left( 1 + \sqrt{\frac{\lambda}{2-\lambda}} \right)  {\rm det} \left[ 1 - \frac{\sqrt{\lambda(2-\lambda)}}{2} {\rm Ai}\left( \frac{x+y}{2} \right) \right]_{L^2(s,\infty)} \nonumber \\
 &+ & \frac{1}{2}\left( 1 - \sqrt{\frac{\lambda}{2-\lambda}}\right) {\rm det} \left[ 1 + \frac{\sqrt{\lambda(2-\lambda)}}{2} {\rm Ai}\left( \frac{x+y}{2} \right) \right]_{L^2(s,\infty)}
\label{S5}
\end{eqnarray}
for $t \gg 1$. The right-hand side is identical to the generating function for GOE (see Eq.~(6.27) in Ref.~\cite{Bornemann_2010}).

\section{Derivation of Eq.~(19)}
The derivation for Eq.~\eqref{G2_dif} of the main text is based on the Taylor expansion for $\cosh(  2 \sqrt{\lambda}  \hat{F}_{  \lfloor 2t + s (2t)^{1/3}/2 \rfloor })$. For brevity, in the following, we use a notation $\hat{A}(t,s) :=  2 \hat{F}_{  \lfloor 2t + s (2t)^{1/3}/2 \rfloor}$. Then, we can show 
\begin{eqnarray}
\cosh \left(\sqrt{\lambda} \hat{A}(t,s) \right) &=& \sum_{n=0}^{\infty} \frac{1}{(2n)!} \left( \sqrt{\lambda} \hat{A}(t,s) \right)^{2n} \nonumber \\
&=& \sum_{n=0}^{\infty} \frac{\lambda^n}{(2n)!} \left( \hat{A}(t,s) \right)^{2n}.
\label{S6}
\end{eqnarray}
Thus, we get
\begin{eqnarray}
\left. \frac{d^n}{d\lambda^n}\cosh \left(\sqrt{\lambda} \hat{A}(t,s) \right)  \right|_{\lambda=0} &=& \frac{n!}{(2n)!}  \left( \hat{A}(t,s) \right)^{2n} \\
&=& \frac{4^n n!}{(2n)!}  \left( \hat{F}_{  \lfloor 2t + s (2t)^{1/3}/2 \rfloor} \right)^{2n}.
\label{S6}
\end{eqnarray}
Equation~\eqref{S6} immediately leads to Eq.~\eqref{G2_dif} of the main text.

\section{Derivation of Eq.~(21)}
Following the same procedure in the derivation of Eq.~\eqref{S6}, we can derive Eq.~\eqref{oddmoments} of the main text. 
What we need is explicit expressions for the Taylor expansion of $\cosh \left( \sqrt{\lambda(2-\lambda)} \hat{A}(t,s) \right)$  and $ \sqrt{\lambda/(2-\lambda)} \sinh \left( \sqrt{\lambda(2-\lambda) } \hat{A}(t,s) \right)$. 
First, we expand  $\cosh \left( \sqrt{\lambda(2-\lambda)} \hat{A}(t,s) \right)$ with respect to $\lambda$:
\begin{eqnarray}
\cosh \left( \sqrt{\lambda(2-\lambda)} \hat{A}(t,s) \right) &=& \sum_{n=0}^{\infty} \frac{ \left( \hat{A}(t,s) \right)^{2n} }{(2n)!} \lambda^n (2-\lambda)^n \nonumber \\
&=& \sum_{n=0}^{\infty} \sum_{k=0}^{n} \frac{ \left( \hat{A}(t,s) \right)^{2n} }{ (2n)! } {}_n C_k 2^{n-k} (-1)^k \lambda^{n+k} \nonumber \\
&=& \sum_{m=0}^{\infty} \sum_{k=0}^{ \lfloor \frac{m}{2} \rfloor} \frac{ (-1)^k 2^{m-2k} {}_{m-k}C_k }{(2m-2k)!} \left( \hat{A}(t,s) \right)^{2m-2k} \lambda^m.
\label{S7}
\end{eqnarray}
Second, expanding $\sinh \left( \sqrt{\lambda(2-\lambda) } \hat{A}(t,s) \right)$ in the same way, we obtain
\begin{eqnarray}
\sqrt{\frac{\lambda}{2-\lambda}} \sinh \left( \sqrt{\lambda(2-\lambda) } \hat{A}(t,s) \right) &=& \sqrt{\frac{\lambda}{2-\lambda}} \sum_{n=0}^{\infty} \frac{ \left( \hat{A}(t,s) \right)^{2n+1} }{ (2n+1)! } \lambda^n(2-\lambda)^n \sqrt{\lambda(2-\lambda)}  \nonumber \\
&=& \sum_{n=0}^{\infty} \frac{ \left( \hat{A}(t,s) \right)^{2n+1} }{ (2n+1)! } \lambda^{n+1} \sum_{k=0}^{n} {}_n C_{k} 2^{n-k} (-\lambda)^k \nonumber \\
&=& \sum_{n=0}^{\infty} \sum_{k=0}^{n} \frac{ \left( \hat{A}(t,s) \right)^{2n+1} }{ (2n+1)! }  {}_nC_k 2^{n-k} (-1)^k \lambda^{n+k+1} \nonumber \\
&=& \sum_{m=1}^{\infty} \sum_{k=0}^{ \lfloor \frac{m-1}{2} \rfloor} \frac{ (-1)^k 2^{m-2k-1} {}_{m-k-1}C_k }{ (2m-2k-1)! } \left( \hat{A}(t,s) \right)^{2m-2k-1} \lambda^m.
\label{S8}
\end{eqnarray}
Finally, differentiating Eqs.~\eqref{S7} and \eqref{S8} with respect to $\lambda$ for $n$ times, we can straightforwardly derive Eq.~\eqref{oddmoments} of the main text.

\clearpage
\section{Dependence on initial states} \label{DIS}
We numerically and analytically investigate dependence of Eqs.~\eqref{evenmoments} and \eqref{oddmoments} in the main text on initial states. First, we show numerical results for 
the first and second moments $M_1(t, l_{t,s})$ and $M_2(t, l_{t,s})$ for the cumulative correlation operator, which are obtained by using three different initial periodic product states being similar to the alternating state. Then, we heuristically find that our results shown in the main text survive in two of these initial states while those disappear in the remaining initial state. Second, we give analytical explanation for this numerical findings by employing the stationary phase method. Finally, we identify a general condition ensuring the emergence of the GOE and GSE behaviors for the initial periodic product states.

\subsection{Numerical investigation}
We numerically study how our result is affected by choices of initial states. For this purpose, we consider the following initial product state defined by
\begin{eqnarray}
\ket{\psi(0)} := \prod_{p=-\infty}^{\infty} \left(  \hat{a}_p^{\dagger} \right)^{S_p} \ket{0}, 
\label{S_D}
\end{eqnarray}
where $S_p~ (p \in \mathbb{Z})$ is an integer taking $0$ or $1$ and determines the initial particle distribution over the lattice. Here, we define $\left(  \hat{a}_p^{\dagger} \right)^0 = 1$. Following the almost same calculation to derive Eq.~\eqref{S_C_con}, we can obtain the two-point correlator with Eq.~\eqref{S_D}:
\begin{eqnarray}
C_{m,n}(t) = {\rm i}^{n-m} \sum_{p=-\infty}^{\infty} S_p J_{m-p}(2t) J_{n-p}(2t).
\label{S_CD}
\end{eqnarray}

\begin{figure}[b]
\begin{center}
\includegraphics[keepaspectratio, width=18cm]{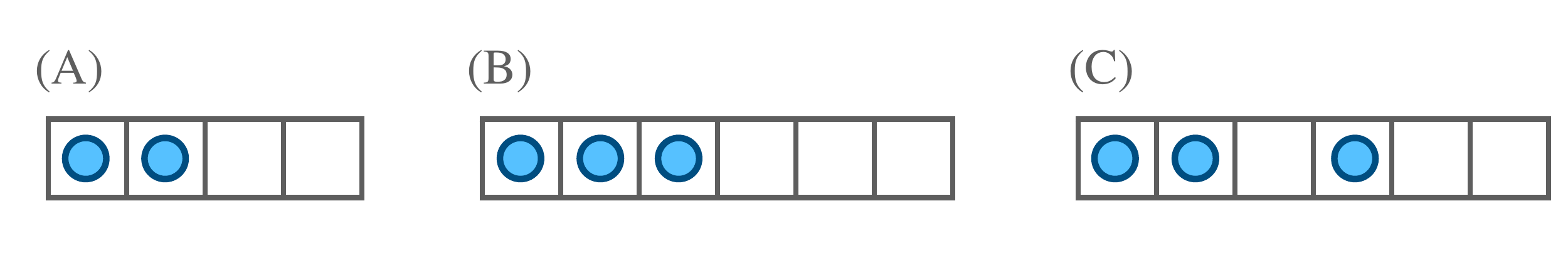}
\caption{
Schematics for the unit cells corresponding to (A), (B), and (C). The label number of the leftmost site is $4p$ for (A) and $6p$ for (B) and (C) with an integer $p$ (see the details of the paragraph just after Eq.~\eqref{S_CD}).
} 
\label{sfig1} 
\end{center}
\end{figure}

We numerically investigate the first and second moments $M_1(t, l_{t,s})$ and $M_2(t, l_{t,s})$ for Eq.~\eqref{S_CD} by changing $S_p$.
The initial distribution $S_p$ addressed here is the following three cases:
 \begin{enumerate}
   \item[(A)] $(S_{4p}, S_{4p+1},S_{4p+2}, S_{4p+3}) = (1,1,0,0) ~ (p \in \mathbb{Z})$, 
   \item[(B)] $(S_{6p}, S_{6p+1},S_{6p+2}, S_{6p+3}, S_{6p+4}, S_{6p+5}) = (1,1,1,0,0,0)~(p \in \mathbb{Z})$, 
   \item[(C)] $(S_{6p}, S_{6p+1},S_{6p+2}, S_{6p+3}, S_{6p+4}, S_{6p+5}) = (1,1,0,1,0,0)~(p \in \mathbb{Z})$. 
\end{enumerate}
The unit cells corresponding to the above are schematically illustrated in Fig.~\ref{sfig1}.

\begin{figure}[b]
\begin{center}
\includegraphics[keepaspectratio, width=18cm]{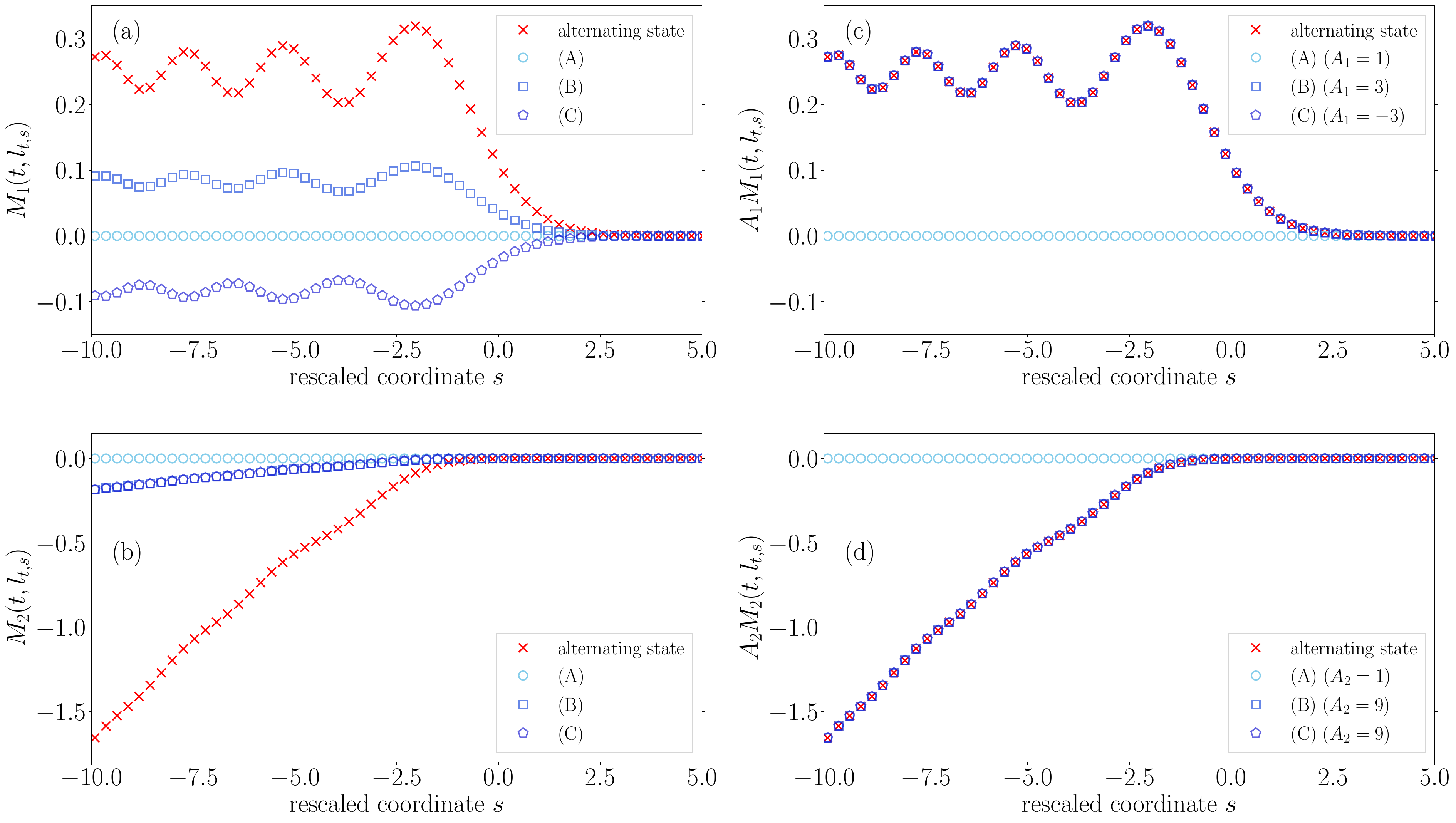}
\caption{
Numerical investigation for the dependence of the first and second moments $M_1(t, l_{t,s})$ and $M_2(t, l_{t,s})$ on the initial states (A), (B), and (C). (a,b) $M_1(t, l_{t,s})$ and $M_2(t, l_{t,s})$ at $t=200$ as a function of a rescaled coordinate $s$ defined through $l_{t,s} = \lfloor 2t + s(2t)^{1/3}/2 \rfloor$. The cross mark, circle, square, and pentagon represent the numerical results with the alternating state, state (A), state (B), and state (C), respectively (see the details of these states just after Eq.~\eqref{S_CD}). (c,d) Rescaled figures corresponding to (a,b). In the initial state (B), we multiply the $y$-axes of (c) and (d) by $A_1=3$ and $A_2=9$, respectively, while in the initial state (C) we multiply the $y$-axes of (c) and (d) by $A_1=-3$ and $A_2=9$, respectively. One can see that the results for the initial states (B) and (C) clearly show excellent agreement with the results for the alternating state after these rescaling of the $y$-axes. On the other hand, the results for the initial state (A) do not match the results for the alternating state as far as we try rescaling. 
} 
\label{sfig2} 
\end{center}
\end{figure}

We numerically compute the moments $M_1(t, l_{t,s})$ and $M_2(t, l_{t,s})$ for these initial product states. Figure~\ref{sfig2} displays the numerical results, from which we find that the initial states (B) and (C) exhibit excellent agreement with the results for the alternating state by adapting appropriate rescaling while the initial state (A) does not. 

\subsection{Analytical investigation for the numerical findings}
We shall give analytical explanation for our numerical results just above. First, we note that the Wick theorem is valid for the initial product state of Eq.~\eqref{S_D}. Then, the generating function for the cumulative correlation operator in the current case is given by Eq.~\eqref{Dform_f} of the main text with the two-point correlator $C_{m,-n}(t)$ of Eq.~\eqref{S_CD}. Thus, the form of $C_{m,-n}(t)$ determines all the moments of the cumulative correlation operator. In other words, it is sufficient to focus only on $C_{m,-n}(t)$ for studying the dependence of our results reported in the main text on initial states. In the following, we investigate the scaling limit of $C_{m,-n}(t)$ in order to investigate the dependence of the first and second moments $M_1(t, l_{t,s})$ and $M_2(t, l_{t,s})$ on the initial states.

Our starting point is the expression of $C_{m,n}(t)$ in the finite system, which is given by Eq.~\eqref{S_C_dis0}. For the readability, we again display it:  
\begin{align}
C_{m,n}(t) = \frac{1}{2L}  \sum_{\alpha=-L}^{L-1} \sum_{\beta=-L}^{L-1} \bra{\psi(0)} {\hat{A}_{\alpha}^{\dagger} \hat{A}_{\beta}} \ket{\psi(0)} \exp \biggl(  {\rm i} 2t \bigl( \cos( \delta k \beta ) - \cos( \delta k \alpha ) \bigl)  + {\rm i} \delta k \bigl(\beta n - \alpha m \bigl) \biggl).
\label{S_C1}
\end{align}
For our initial state of Eq.~\eqref{S_D}, the explicit expression of $ \bra{\psi(0)} {\hat{A}_{\alpha}^{\dagger} \hat{A}_{\beta}} \ket{\psi(0)} $ is given by
\begin{align}
\bra{\psi(0)} {\hat{A}_{\alpha}^{\dagger} \hat{A}_{\beta}} \ket{\psi(0)} &= \frac{1}{2L} \sum_{m=-L}^{L-1} \sum_{n=-L}^{L-1} \bra{\psi(0)} {\hat{a}_{m}^{\dagger} \hat{a}_{n}} \ket{\psi(0)} \exp \biggl( {\rm i} \delta k \bigl( \alpha m  - \beta n \bigl) \biggl)  \\
&= \frac{1}{2L} \sum_{m=-L}^{L-1}  S_m \exp \biggl( {\rm i} \delta k m \bigl( \alpha   - \beta  \bigl) \biggl).
\label{S_C2}
\end{align}
In what follows, we analytically derive the concrete expression of $C_{m,n}(t)$ for the initial states (A), (B), and (C) using Eqs.~\eqref{S_C1} and \eqref{S_C2}.

\subsubsection{Initial state (A)}
We first evaluate $\bra{\psi(0)} {\hat{A}_{\alpha}^{\dagger} \hat{A}_{\beta}} \ket{\psi(0)}$. We assume $L = 2N$ with an even integer $N$.
Using Eq.~\eqref{S_C2} and $S_j$ corresponding to the initial state (A), we obtain
\begin{align}
&\bra{\psi(0)} {\hat{A}_{\alpha}^{\dagger} \hat{A}_{\beta}} \ket{\psi(0)} \\
&= \frac{1}{4N} \sum_{m=-N/2}^{N/2-1}   \Biggl( \exp \biggl[ {\rm i} \delta k (4 m) \bigl( \alpha   - \beta  \bigl) \biggl] + \exp \biggl[ {\rm i} \delta k (4 m + 1) \bigl( \alpha   - \beta  \bigl) \biggl] \Biggl) \\
& = \frac{1}{4N} \Biggl( 1 + \exp \biggl[ {\rm i} \delta k \bigl( \alpha   - \beta  \bigl) \biggl] \Biggl)  \sum_{m=-N/2}^{N/2-1}  \exp \biggl[ {\rm i} \frac{2 \pi}{N} m \bigl( \alpha   - \beta  \bigl) \biggl] \\
& = \frac{1}{4} \Biggl( 1 + \exp \biggl[ {\rm i} \frac{\pi}{2N} \bigl( \alpha   - \beta  \bigl) \biggl] \Biggl) \biggl(  \delta_{\alpha, \beta-3N} + \delta_{\alpha, \beta-2N} + \delta_{\alpha, \beta-N} + \delta_{\alpha, \beta} + \delta_{\alpha, \beta+N} + \delta_{\alpha, \beta + 2N} + \delta_{\alpha, \beta+3N}    \biggl) \label{S_C2A1} \\
& =  \frac{1+ {\rm i} }{4}  \delta_{\alpha, \beta-3L/2} + \frac{1- {\rm i} }{4}  \delta_{\alpha, \beta-L/2} + \frac{1}{2} \delta_{\alpha, \beta} + \frac{1+ {\rm i} }{4}  \delta_{\alpha, \beta+L/2} + \frac{1- {\rm i} }{4}  \delta_{\alpha, \beta+3L/2}.
\label{S_C2A}
\end{align}
Substituting Eq.~\eqref{S_C2A} into Eq.~\eqref{S_C1}, we get
\begin{align}
C_{m,n}(t) = \frac{1}{2} \delta_{m,n} &+ \frac{1- {\rm i} }{8L}  \sum_{\alpha=-L}^{L-1}  \exp \biggl[  {\rm i} 2t \biggl( \cos \bigl( \delta k \alpha + \pi/2\bigl) - \cos \bigl( \delta k \alpha \bigl) \biggl)  + {\rm i} \delta k \alpha \bigl( n -  m \bigl) + {\rm i} \pi n/2 \biggl]  \nonumber  \\
&+ \frac{1+ {\rm i} }{8L}  \sum_{\alpha=-L}^{L-1}  \exp \biggl[  {\rm i} 2t \biggl( \cos \bigl( \delta k \alpha - \pi/2\bigl) - \cos \bigl( \delta k \alpha \bigl) \biggl)  + {\rm i} \delta k \alpha \bigl( n - m \bigl) - {\rm i} \pi n/2 \biggl].
\label{S_C2A_1}
\end{align}
In the thermodynamic limit, this becomes
\begin{align}
C_{m,n}(t) &= \frac{1}{2} \delta_{m,n} + \frac{1 }{ 8 \pi }(1-  {\rm i}){\rm i}^n  \int_{-\pi}^{\pi} d \theta  \exp \biggl[  {\rm i} 2t \biggl( \cos \bigl( \theta + \pi/2\bigl) - \cos \bigl( \theta \bigl) \biggl)  + {\rm i} \theta \bigl( n -  m \bigl)  \biggl]  \nonumber  \\
&~~~~~~~~~~~~+ \frac{1 }{ 8 \pi }(1+  {\rm i})(- {\rm i})^n  \int_{-\pi}^{\pi} d \theta  \exp \biggl[  {\rm i} 2t \biggl( \cos \bigl( \theta - \pi/2\bigl) - \cos \bigl( \theta \bigl) \biggl)  + {\rm i} \theta \bigl( n - m \bigl)  \biggl] \\
&= \frac{1}{2} \delta_{m,n} + \frac{1 }{ 8 \pi }(1-  {\rm i}){\rm i}^n  \int_{-\pi}^{\pi} d \theta  \exp \biggl[   - {\rm i} 2 \sqrt{2}t\sin \bigl( \theta + \pi/4\bigl)   + {\rm i} \theta \bigl( n -  m \bigl)  \biggl]  \nonumber \\
&~~~~~~~~~~~~+ \frac{1 }{ 8 \pi }(1+  {\rm i})(- {\rm i})^n  \underbrace{\int_{-\pi}^{\pi} d \theta  \exp \biggl[  {\rm i}  2 \sqrt{2}t \sin \bigl( \theta - \pi/4\bigl)  + {\rm i} \theta \bigl( n - m \bigl)  \biggl]}_{=: I_{m-n}(t)}.
\label{S_C2A_2}
\end{align}

The important ingredient for the emergence of the GOE and GSE behaviors is the expression of $C_{m,-n}(t)$ under the scaling limit which is taken with $n= \lfloor 2 t + (2 t)^{1/3} x/2  \rfloor$ and $m= \lfloor 2 t + (2 t)^{1/3} y/2  \rfloor$. Using the same stationary phase method used in Sec. II of this supplemental material, we can find that the integrals for the second and third terms on the right-hand side of Eq.~\eqref{S_C2A_2} do not asymptotically approach the Airy function under the scaling limit of the main text. We now show this fact by considering $I_{m-n}(t)$ defined in Eq.~\eqref{S_C2A_2}. In order to consider $C_{m,-n}(t)$, we focus on the behavior of $I_{m+n}(t)$ with $n= \lfloor 2 t + (2 t)^{1/3} x/2  \rfloor$ and $m= \lfloor 2 t + (2 t)^{1/3} y/2  \rfloor$. Then, we obtain
\begin{align}
I_{m+n}(t) \simeq \int_{-\pi}^{\pi} d \theta  \exp \biggl[  {\rm i} t \biggl( 2 \sqrt{2} \sin \bigl( \theta - \pi/4\bigl)  -  4 \theta \biggl) - {\rm i} \theta (2t)^{1/3} \frac{ x+y }{2}    \biggl], 
\label{S_C2A_3}
\end{align}
from which it follows that the function $g(\theta) :=  2 \sqrt{2} \sin ( \theta - \pi/4 )  -  4 \theta $ can induce rapid oscillation of the integrand for $t \gg 1$. One, however, can easily show that $g(\theta)$ does not have an extremum.
Hence, the value of $I_{m+n}(t) $ under the scaling limit cannot be large. The same thing holds for the second term on the right-hand side of Eq.~\eqref{S_C2A_2}. Our finding here implies that $C_{m,-n}(t)$ under the scaling limit becomes very small. This explains our numerical results in Figs.~\ref{sfig2} (a) and (b), where the values of the first and second moments are almost zero.

The mathematical mechanism for the absence of the Airy function under the scaling limit stems from the fact that the contribution from the terms with $\delta_{\alpha, \beta \pm 2N} = \delta_{\alpha, \beta \pm L}$ in Eq.~\eqref{S_C2A1} vanishes because of $e^{{\rm i}\pi} = -1$. As noted just after Eq.~\eqref{S_bessel}, these terms are responsible for the emergence of the Bessel function of the first kind leading to the Airy function, and thus the absence of such terms explains the disappearance of the GOE and GSE behaviors in Fig.~\ref{sfig2}.

\subsubsection{Initial state (B)}
We evaluate $\bra{\psi(0)} {\hat{A}_{\alpha}^{\dagger} \hat{A}_{\beta}} \ket{\psi(0)}$. We assume $L = 3N$ with an even integer $N$.
Using Eq.~\eqref{S_C2} and $S_j$ corresponding to the initial state (B), we obtain
\begin{align}
&\bra{\psi(0)} {\hat{A}_{\alpha}^{\dagger} \hat{A}_{\beta}} \ket{\psi(0)} \\
&= \frac{1}{6N} \sum_{m=-N/2}^{N/2-1}   \Biggl( \exp \biggl[ {\rm i} \delta k (6 m) \bigl( \alpha   - \beta  \bigl) \biggl] + \exp \biggl[ {\rm i} \delta k (6 m + 1) \bigl( \alpha   - \beta  \bigl) \biggl] + \exp \biggl[ {\rm i} \delta k (6 m + 2) \bigl( \alpha   - \beta  \bigl) \biggl] \Biggl)  \\
& = \frac{1}{6N} \Biggl( 1 + \exp \biggl[ {\rm i} \delta k \bigl( \alpha   - \beta  \bigl) \biggl] + \exp \biggl[ {\rm i} 2 \delta k \bigl( \alpha   - \beta  \bigl) \biggl]  \Biggl)  \sum_{m=-N/2}^{N/2-1}  \exp \biggl[ {\rm i} \frac{2 \pi}{N} m \bigl( \alpha   - \beta  \bigl) \biggl] \\
& = \frac{1}{6} \Biggl( 1 + \exp \biggl[ {\rm i} \frac{\pi}{3N} \bigl( \alpha   - \beta  \bigl) \biggl] + \exp \biggl[ {\rm i} \frac{2\pi}{3N} \bigl( \alpha   - \beta  \bigl) \biggl]  \Biggl)  \nonumber  \\
&~~~\times \biggl( \delta_{\alpha, \beta-5N} + \delta_{\alpha, \beta-4N} + \delta_{\alpha, \beta-3N} + \delta_{\alpha, \beta-2N} + \delta_{\alpha, \beta-N} + \delta_{\alpha, \beta} + \delta_{\alpha, \beta+N} + \delta_{\alpha, \beta + 2N} + \delta_{\alpha, \beta + 3N} + \delta_{\alpha, \beta + 4N} + \delta_{\alpha, \beta + 5N}     \biggl)  \\
& =  \frac{1}{6}  \delta_{\alpha, \beta-L} +  \frac{1}{2} \delta_{\alpha,\beta} + \frac{1}{6} \delta_{\alpha, \beta+L} + {\rm (other~terms)}.
\label{S_C2B}
\end{align}
The terms with $\delta_{\alpha, \beta \pm L}$ in Eq.~\eqref{S_C2B} play the same role in the first and third terms of Eq.~\eqref{S_A}, and we can derive 
\begin{align}
C_{m,n}(t) = \frac{1}{2} \delta_{m,n} + \frac{1}{6} {\rm i}^{n+m} J_{n-m}(4t) + {\rm (other~terms)}
\label{S_C2B_1}
\end{align}
in the thermodynamic limit. The other terms become small under the scaling limit because of the absence of extremum mentioned just after Eq.~\eqref{S_C2A_3}. 
The two-point correlator $C_{m,n}(t)$ of Eq.~\eqref{S_C2B_1} is similar to that of Eq.~\eqref{S_C_con}, but the factors for the Bessel functions of the first kind are different. This is the reason why we need to multiply the $y$-axes by $3$ and $9$ in Figs.~\ref{sfig2}(c) and (d), respectively.

\subsubsection{Initial state (C)}
We evaluate $\bra{\psi(0)} {\hat{A}_{\alpha}^{\dagger} \hat{A}_{\beta}} \ket{\psi(0)}$. We assume $L = 3N$ with an even integer $N$.
Using Eq.~\eqref{S_C2} and $S_j$ corresponding to the initial state (C), we obtain
\begin{align}
&\bra{\psi(0)} {\hat{A}_{\alpha}^{\dagger} \hat{A}_{\beta}} \ket{\psi(0)} \\
&= \frac{1}{6N} \sum_{m=-N/2}^{N/2-1}   \Biggl( \exp \biggl[ {\rm i} \delta k (6 m) \bigl( \alpha   - \beta  \bigl) \biggl] + \exp \biggl[ {\rm i} \delta k (6 m + 1) \bigl( \alpha   - \beta  \bigl) \biggl] + \exp \biggl[ {\rm i} \delta k (6 m + 3) \bigl( \alpha   - \beta  \bigl) \biggl] \Biggl)  \\
& = \frac{1}{6N} \Biggl( 1 + \exp \biggl[ {\rm i} \delta k \bigl( \alpha   - \beta  \bigl) \biggl] + \exp \biggl[ {\rm i} 3 \delta k \bigl( \alpha   - \beta  \bigl) \biggl]  \Biggl)  \sum_{m=-N/2}^{N/2-1}  \exp \biggl[ {\rm i} \frac{2 \pi}{N} m \bigl( \alpha   - \beta  \bigl) \biggl] \\
& = \frac{1}{6} \Biggl( 1 + \exp \biggl[ {\rm i} \frac{\pi}{3N} \bigl( \alpha   - \beta  \bigl) \biggl] + \exp \biggl[ {\rm i} \frac{\pi}{N} \bigl( \alpha   - \beta  \bigl) \biggl]  \Biggl)  \nonumber  \\
&~~~\times \biggl( \delta_{\alpha, \beta-5N} + \delta_{\alpha, \beta-4N} + \delta_{\alpha, \beta-3N} + \delta_{\alpha, \beta-2N} + \delta_{\alpha, \beta-N} + \delta_{\alpha, \beta} + \delta_{\alpha, \beta+N} + \delta_{\alpha, \beta + 2N} + \delta_{\alpha, \beta + 3N} + \delta_{\alpha, \beta + 4N} + \delta_{\alpha, \beta + 5N}     \biggl) \\
& =  -\frac{1}{6} \delta_{\alpha, \beta-L} +  \frac{1}{2} \delta_{\alpha,\beta} - \frac{1}{6} \delta_{\alpha, \beta+L} + {\rm (other~terms)}.
\label{S_C2C}
\end{align}
In the same way as the initial state (B), the terms with $\delta_{\alpha, \beta \pm L}$ in Eq.~\eqref{S_C2C} leads to the Bessel function of the first kind, and we eventually obtain 
\begin{align}
C_{m,n}(t) = \frac{1}{2} \delta_{m,n} - \frac{1}{6} {\rm i}^{n+m} J_{n-m}(4t) + {\rm (other~terms)}
\end{align}
in the thermodynamic limit. The other terms become small under the scaling limit because of the same reason for the initial state (B). 
Note that the factor for the Bessel function of the first kind is different from that of Eq.~\eqref{S_C_con}. This explains the reason why we need to multiply the $y$-axes by $-3$ and $9$ in Figs.~\ref{sfig2}(c) and (d), respectively.

\subsection{Condition for the emergence of the GOE and GSE behaviors in the initial periodic product state (S-36)}
\begin{figure}[b]
\begin{center}
\includegraphics[keepaspectratio, width=18cm]{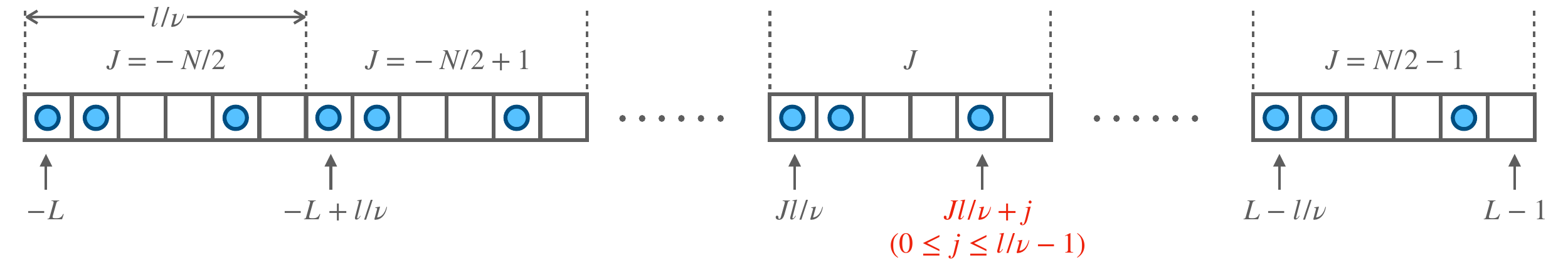}
\caption{
Schematic configuration of the initial state used in Sec.~VIII~C. The size of the unit cell is $l/\nu$ with an integer $l$ and the total number of the lattice points is $2L = N l/\nu$ with an even number $N$. The vertical dashed lines display boundaries between the unit cells, and the filled circles denote the fermions. The integers $J$ and $j$ in the figure are labels for the unit cells and the sites inside each unit cell. Using them, we can specify all the sites by $J l/\nu + j$ for $ -N/2 \leq J \leq N/2-1$ and $0 \leq j \leq l/\nu-1$, as depicted in the figure.
} 
\label{sfig3} 
\end{center}
\end{figure}

We derive a condition for the emergence of the GOE and GSE behaviors in dynamics starting from the $l/\nu$-periodic product state defined by Eq.~\eqref{S_D} with $S_{j} = S_{j+l/\nu}$ and an integer $l$. Here, we consider a filling factor $\nu$ for which $1/\nu$ is an integer. Figure~\ref{sfig3} schematically illustrates the initial state, where a lattice-point number in a unit cell is $l/\nu$. We suppose that the total lattice number is given by $2L = N l/\nu $ with an even integer $N$ and the fermions occupy $l$ sites in each unit cell. Under this setup, a site is labeled by $ J l/\nu + j$ for $ -N/2 \leq J \leq N/2-1$ and $0 \leq j \leq l/\nu-1$. Here, the integers $J$ and $j$ are labels for the unit cells and the sites in each unit cell, respectively (see Fig.~\ref{sfig3}). 

We compute $\bra{\psi(0)} {\hat{A}_{\alpha}^{\dagger} \hat{A}_{\beta}} \ket{\psi(0)}$ with Eq.~\eqref{S_D} using the site labels introduced just above.
Employing Eq.~\eqref{S_C2}, we obtain  
\begin{align}
\bra{\psi(0)} {\hat{A}_{\alpha}^{\dagger} \hat{A}_{\beta}} \ket{\psi(0)} 
& =  \frac{\nu}{lN} \sum_{m=-L}^{L-1} S_m  \exp \biggl[ {\rm i} \delta k  m ( \alpha   - \beta  \bigl) \biggl] \\
& =  \frac{\nu}{lN} \sum_{J=-N/2}^{N/2-1} \sum_{j=0}^{ l/\nu-1 } \underbrace{S_{J l/\nu+j}}_{=S_j}   \exp \biggl[ {\rm i} \delta k  \left( \frac{J l}{\nu}+j \right) ( \alpha   - \beta  \bigl) \biggl] \\
& = \frac{\nu}{lN} \left( \sum_{j=0}^{ l/\nu-1 } S_j \exp \biggl[ {\rm i} \delta k j \bigl( \alpha   - \beta  \bigl) \biggl] \right) \left( \sum_{J=-N/2}^{N/2-1}  \exp \biggl[ {\rm i} \delta k \left( \frac{J l}{\nu} \right) \bigl( \alpha   - \beta  \bigl) \biggl]  \right) \\
& = \frac{\nu}{l} \left( \sum_{j=0}^{ l/\nu-1 } S_j \exp \biggl[ {\rm i} \delta k j \bigl( \alpha   - \beta  \bigl) \biggl] \right)  \left( \sum_{m= -l/\nu+1 }^{  l/\nu-1 } \delta_{\alpha, \beta + mN}  \right) \\
& = \frac{\nu}{l} \left( \sum_{j=0}^{ l/\nu-1 } S_j \exp \biggl[ {\rm i} \delta k j \bigl( \alpha   - \beta  \bigl) \biggl] \right)  \left(  \delta_{\alpha, \beta - L} + \delta_{\alpha, \beta + L} +  \sum_{  \substack{m= -l/\nu+1 \\ (m\neq \pm l/(2\nu) )}    }^{l/\nu-1}  \delta_{\alpha, \beta + mN}  \right),
\label{S_general1}
\end{align}
where we use the site label $m = J l/\nu + j$ in the second line and the $l/\nu$-periodicity of $S_j$ is used in the third line. As shown in Sec.~VIII B, the terms with $\delta_{\alpha, \beta \pm L}$ of Eq.~\eqref{S_general1} give rise to the Bessel function of the first kind leading to the Airy function. Hence, we identify a condition for the emergence of the GOE and GSE behaviors as 
\begin{align}
     &\sum_{j=0}^{ l/\nu-1 } S_j \exp \biggl[   {\rm i} \delta k j (\alpha-\beta)  \biggl] \delta_{\alpha,\beta \pm L} \neq 0 \\
\iff &\sum_{j=0}^{ l/\nu-1 } S_j \exp \biggl[  \pm {\rm i} L \delta k j  \biggl] \neq 0 \\
\iff &\sum_{j=0}^{ l/\nu-1 } S_j (-1)^j \neq 0, 
\label{S_general2}
\end{align}
where we use $\delta k L = \pi$ in the third line. The condition~\eqref{S_general2} always holds for odd $l$, while it is not necessarily satisfied for even $l$. Thus, the number of the initial state exhibiting the GOE and GSE behaviors is larger than that of states not exhibiting them.

\end{document}